\documentclass[aps,prb,twocolumn,floatfix,notitlepage, superscriptaddress]{revtex4-2}
\usepackage[utf8]{inputenc}
\usepackage{amsfonts,amsmath,amssymb,graphicx,epstopdf}
\usepackage{xcolor}
\usepackage{natbib}
\usepackage[colorlinks]{hyperref}
\definecolor{darkred}{rgb}{0.5,0,0}
\definecolor{darkgreen}{rgb}{0,0.5,0}
\definecolor{darkblue}{rgb}{0,0,0.5}
\hypersetup{colorlinks,
linkcolor=darkred,
filecolor=darkgreen,
urlcolor=darkblue,
citecolor=darkgreen}
\usepackage{braket}
\usepackage{overpic}

\newcommand{\nep}{\operatorname{e}}

\newcommand{\up}{\uparrow}
\newcommand{\down}{\downarrow}

\usepackage{orcidlink}

\begin{document}
\title{Nonstabilizerness in the unitary and monitored quantum dynamics of XXZ-staggered and SYK models}

\author{Angelo Russomanno~\orcidlink{0009-0000-1923-370X}}
\affiliation{Dipartimento di Fisica ``E. Pancini'', Universit\`a di Napoli Federico II, Complesso di Monte S. Angelo, via Cinthia, I-80126 Napoli, Italy}

\author{Gianluca Passarelli~\orcidlink{0000-0002-3292-0034}}
\affiliation{Dipartimento di Fisica ``E. Pancini'', Universit\`a di Napoli Federico II, Complesso di Monte S. Angelo, via Cinthia, I-80126 Napoli, Italy}

\author{Davide Rossini~\orcidlink{0000-0002-9222-1913}}
\affiliation{Dipartimento di Fisica dell’Università di Pisa and INFN, 
Largo Pontecorvo 3, I-56127 Pisa, Italy}

\author{Procolo Lucignano~\orcidlink{0000-0003-2784-8485}}
\affiliation{Dipartimento di Fisica ``E. Pancini'', Universit\`a di Napoli Federico II, Complesso di Monte S. Angelo, via Cinthia, I-80126 Napoli, Italy}

\begin{abstract}
We consider the quantum-state-diffusion dynamics of the XXZ-staggered spin chain, also focusing on its noninteracting {XX-staggered} limit, and of the Sachdev-Ye-Kitaev (SYK) model. We {describe the process through quantum trajectories} and evaluate the nonstabilizerness (also known as ``magic'') along the trajectories, quantified through the stabilizer R\'enyi entropy (SRE). 
In the absence of measurements, we find that the SYK model is the only one in which the time-averaged SRE saturates the random state bound and has a scaling with the system size that is well described by the theoretical prediction for quantum chaotic systems. In the presence of measurements, we {numerically} find that the {{steady-state}} SRE versus the coupling strength to the environment is well fitted by a {generalized Lorentzian} function. The scaling of the fitting parameters with the system size suggests that the {{steady-state} SRE} {linearly} increases with the {system} size in all the considered cases, {and displays} no measurement-induced quantum transition, {as confirmed by the curves of the steady-state SRE versus the system size}.
\end{abstract}

\maketitle
\section{Introduction}
Entanglement is an important measure of quantum complexity and a relevant {necessary ingredient for quantum advantage, but not a sufficient one}~\cite{Nielsen}. It is so crucial that the recent discovery of a different, yet independent, measure of quantum complexity came as a surprise. This quantity is called nonstabilizerness (or ``magic'') and quantifies {the distance of} a pure state from the set of stabilizer states~\cite{veitch2014,howard2014-2,bravyi2016,bravyi2016-2,bravyi2019,heinrich2019,wang2019,wang2020,heimendahl2021,jiang2023,haug2023-3}, which can be prepared applying Clifford quantum circuits. Such circuits have the remarkable property of being efficiently simulatable on a classical computer, despite being able to generate entanglement~\cite{Fisher_2023}. {Although the Gottesman-Knill theorem~\cite{gottesman1998,gottesman1998-2} relating stabilizer states to Clifford quantum circuits is textbook knowledge, only recently a manageable measure of distance from the set of classically-simulatable stabilizer states, i.e., the stabilizer R\'enyi entropy (SRE), was introduced~\cite{leone2022,oliviero2022}.} Such a discovery recently fostered excitement in the scientific community, as witnessed by the flourishing amount of publications on this subject~\cite{bravyi2005, howard2014, veitch2014, chitambar2019, seddon2019, zhou2020, liu2022, PhysRevA.107.022429, rattacaso2023, Odavic_2023, Leone_2024, tirrito2023, paviglianiti2024estimatingnonstabilizernessdynamicssimulating,dowling2024magicheisenbergpicture,PhysRevB.111.054301,Turkeshi_2025,PhysRevA.110.022436,passarelli2024chaosmagicdissipativequantum,niroula2024phasetransitionmagicrandom,PhysRevB.110.045101,PhysRevB.110.045101,odavi2025stabilizerentropynonintegrablequantum,passarelli2025nonstabilizernessboundarytimecrystal}.

Another research avenue dealing with quantum complexity that has emerged in the last few years is the one of entanglement transitions in monitored systems. A monitored system is a quantum system that evolves under the effect of {continuous measurements} from an environment, that can be discrete as in the case of quantum circuits~\cite{Fisher_2023}, or continuous as for quantum trajectories~\cite{Plenio,fazio2024manybodyopenquantumsystems,Daley2014}. 
Quite importantly, the state along each trajectory is a pure one, and one gets the usual description in terms of density matrix evolving under a Lindblad equation only on average. On this pure state, it is possible to evaluate the entanglement (usually by means of the half-system entanglement entropy) and to study its behavior, after averaging over trajectories in the {{steady-state} regime}. 

It has been found that the interplay between the entangling effect of the unitary part of the dynamics and the disentangling role of {continuous measurements} can give rise to different dynamical regimes where the entanglement entropy in the {steady-state} regime scales differently with the system size. These measurement-induced entanglement transitions have been found in several contexts ranging from quantum 
circuits~\cite{Li2018, Chan2019, Skinner2019, Szyniszewski2019, Vasseur2021, Bao2021, Nahum2020, Chen2020,Li2019, Jian2020, Li2021, Szyniszewski2020, Turkeshi2020, Lunt2021, Sierant2022_B, Nahum2021, Zabalo2020, Sierant2022_A, Chiriaco2023, Klocke2023,lirasolanilla2024},
to integrable or solvable~\cite{PhysRevResearch.6.043246,chahine2023entanglement,delmonte2024,karevski2024,Passarelli_2024,DeLuca2019,Nahum2020, Buchhold2021,Jian2022, Coppola2022, Fava2023, Poboiko2023, Jian2023, Merritt2023, Alberton2021, Turkeshi2021, Szyniszewski2022, Turkeshi2022, Piccitto2022, Piccitto2022e, Tirrito2022, Paviglianiti2023, Lang2020, Minato2022, Zerba2023, paviglianiti2023enhanced,chatterjee2024, gda_EPJB, Le_Gal_2024,Russomanno2023_longrange} and nonintegrable~\cite{Lunt2020,Rossini2020, Tang2020, Fuji2020, Sierant2021, Doggen2022, Altland2022, li2024monitored} monitored Hamiltonian systems, and even in cases of measurement-only dynamics of nonlocal strings~\cite{Ippoliti2021, Sriram2022,Piccitto2023}, and in the average density matrix of systems with power-law Lindbladians~\cite{PhysRevB.109.214204}.

One could consider the behavior of the nonstabilizerness in monitored systems as a different measure of quantum complexity. This analysis has been actually performed for a Clifford quantum circuit to which $T$ gates creating {nonstabilizerness} and random projective measurements were applied, and a transition both in entanglement and {nonstabilizerness} was found~\cite{PRXQuantum.5.030332}. In Ref.~\cite{PhysRevResearch.6.L042030} it was argued that the entanglement transition occurs at a different point than the {nonstabilizerness} transition, so the question is still debated. Moreover, a transition in the behavior of the {nonstabilizerness} has also been observed in a measurement-only circuit~\cite{tarabunga2024magictransitionmeasurementonlycircuits}.

In this contribution we follow this line of research and study two many-body quantum systems---namely, the {XXZ-staggered spin-1/2 chain}~\cite{xing2023interactions} and the Sachdev-Ye-Kitaev (SYK) model~\cite{Sachdev_1993, Kitaev_2015}---in contact with an environment performing continuous measurements that, on average, provide a Lindblad evolution. We choose a specific measurement scheme, corresponding to the so-called quantum-state-diffusion (QSD) unraveling. Averaging over the trajectories generated by this process gives rise to the Lindblad equation for observables. We consider the nonstabilizerness averaged over trajectories, quantified through the SRE, for which the Lindblad approach cannot be applied, because this quantity is nonlinear in the quantum state. 
{The dynamics of both models is restricted to the subspace with vanishing total $z$ magnetization. We first focus on the time behavior of the nonstabilizerness in the unitary limit, where no measurements are performed, studying how the time-averaged SRE depends on the system size.
Then, we switch on the coupling $\gamma$ with the monitoring environment and discuss the behavior of the SRE with $\gamma$ and the size.}

{For the unitary dynamics of the SYK model, we find a fast increase with the system size, which, after {an initial transient behavior}, attains the behavior of a random state ({random-phase state}s and Haar random states provide the same result). 
We believe this to be intimately connected with the fact that the SYK model is the most chaotic quantum model, exhibiting fast scrambling~\cite{PhysRevD.94.106002,Roberts_2018}, a nonzero entropy density at vanishing temperature~\cite{PhysRevB.94.035135}, and a volume-law bipartite entanglement entropy for all the eigenstates (even for the ground state)~\cite{Balents_2018,PhysRevD.100.041901}. Thus, the long-time dynamics of this model coincides with that of a random state, also when considering a very nonlocal quantity as the SRE. Moreover, the time-averaged SRE, both in the SYK model and in the {random-phase state}, increases as the logarithm of the dimension of the relevant Hilbert subspace, in agreement with the theoretical predictions for a generic quantum chaotic system~\cite{leone2021}.}
{On the opposite hand, the time-averaged SRE in the XXZ-staggered model shows a much slower increase with the system size, although still approximately linear, both in the integrable XX-staggered limit and in the generic nonintegrable case. We argue that this is due to the fact that the XXZ model has local conserved quantities, while in the SYK model the Hamiltonian is a conserved, but nonlocal, quantity. A similar result has been reported in Ref.~\cite{tirrito2024anticoncentrationmagicspreadingergodic}, where relaxation in time towards the Haar random state value could occur only if the conservation of the local Hamiltonian was broken by applying a periodic driving.}

{Switching on $\gamma$, we analyze the behavior of the SRE averaged over trajectories and find that, after an initial transient, it reaches a {steady-state} value. {For all the considered models w}e {find} that the dependence of the {steady-state} SRE on $\gamma$ is well fitted by a generalized Lorentzian function. Studying how the fitting parameters depend on the system size $L$, we infer that the {{steady-state}} SRE appears to be always linear in $L$, {as confirmed by the curves of the steady-state SRE versus the system size}. Thus, even in the presence of an external monitoring, the state of the system cannot be described by a Clifford circuit, while more and more $T$ gates are needed to describe it as the system size is increased.}

{The $\gamma\to 0$ limit of the steady-state SRE in the nonintegrable XXZ-staggered model coincides with the random-state value, {in contrast with} the time average in the unitary case, {that lies well below the random-state value}. Therefore the $\gamma\to 0$ limit is singular, {because it does not coincide with the corresponding value in the unitary case}. This follows from the fact that a small amount of {stochastic noise due to quantum measurements} breaks the conservation of energy, that is the only {local} integral of motion of such model restricted to the subspace of interest, and so relaxation to the random-state value is hindered. For the SYK model -- where the Hamiltonian is nonlocal and there are no local integrals of motion in the Hilbert subspace of interest -- the $\gamma\to 0$ limit {of the steady-state SRE} coincides with the unitary result and the random-state value. On the other hand, the $\gamma\to 0$ limit of the integrable XX-staggered model is smaller than that of the random-state value, because this latter case is equivalent to a quadratic model of free fermions, also when monitoring is applied.

The paper is organized as follows. In Sec.~\ref{mod:sec} we introduce the models considered in the following and define the SRE.  In Sec.~\ref{unidyn:sec} we show our numerical results for the unitary dynamics in the absence of monitoring. In Sec.~\ref{monidyn:sec} we introduce the QSD unraveling with which we describe the measurements from the environment and show the results for the {{steady-state}} SRE in this monitored case. Our conclusions are drawn in Sec.~\ref{conc:sec}. Appendix~\ref{nonstabilizerness:sec} provides details on the SRE (i.e., the quantity we adopt to quantify nonstabilizerness) and describes the numerical method used to evaluate it.
\section{Models and evaluated quantity}\label{mod:sec}

\begin{figure*}[t]
  \begin{center}
   \begin{tabular}{ccc}
   (a) XX-staggered& (b) XXZ-staggered & (c) SYK\\
    \includegraphics[width=58mm]{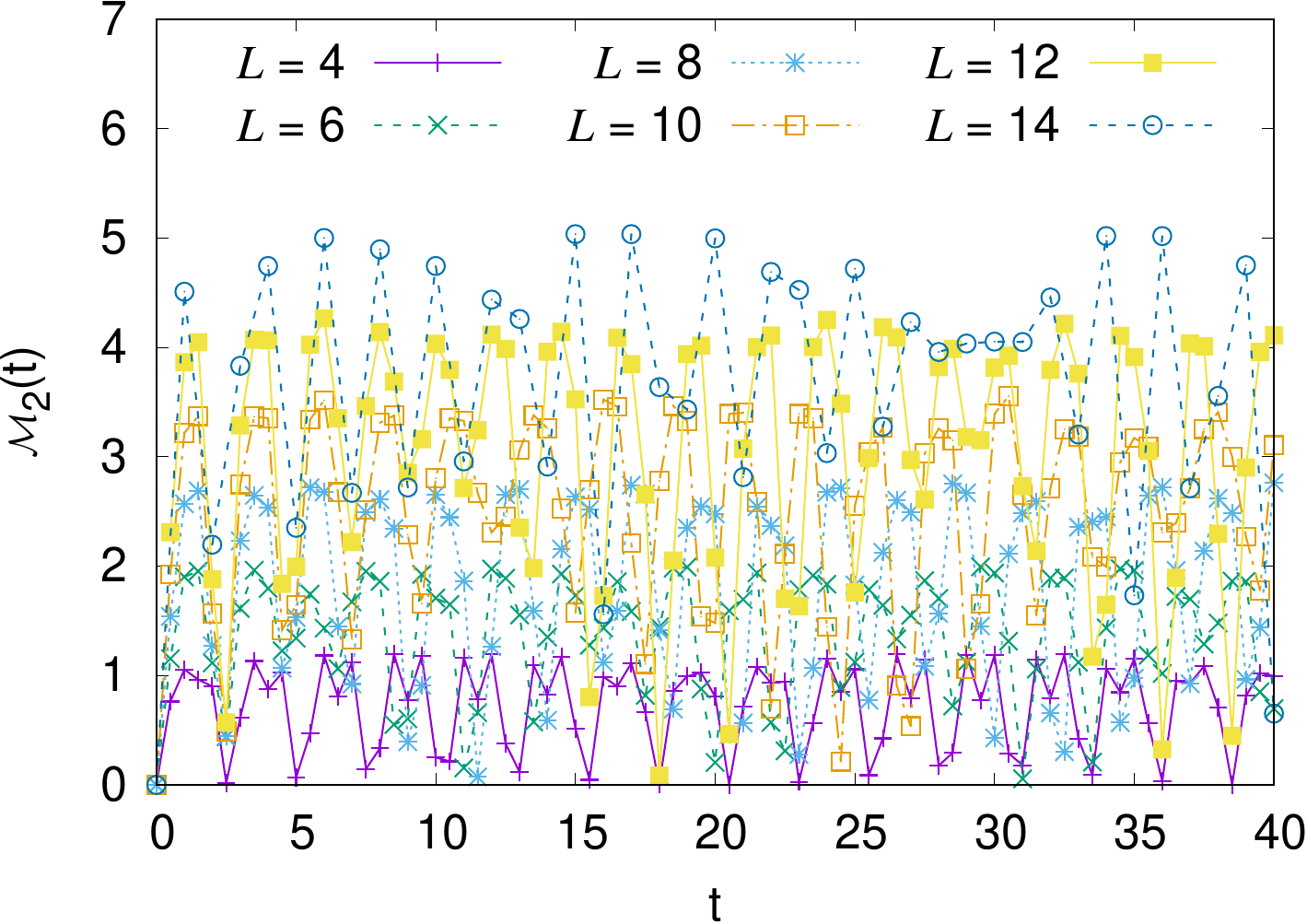}&
    \includegraphics[width=58mm]{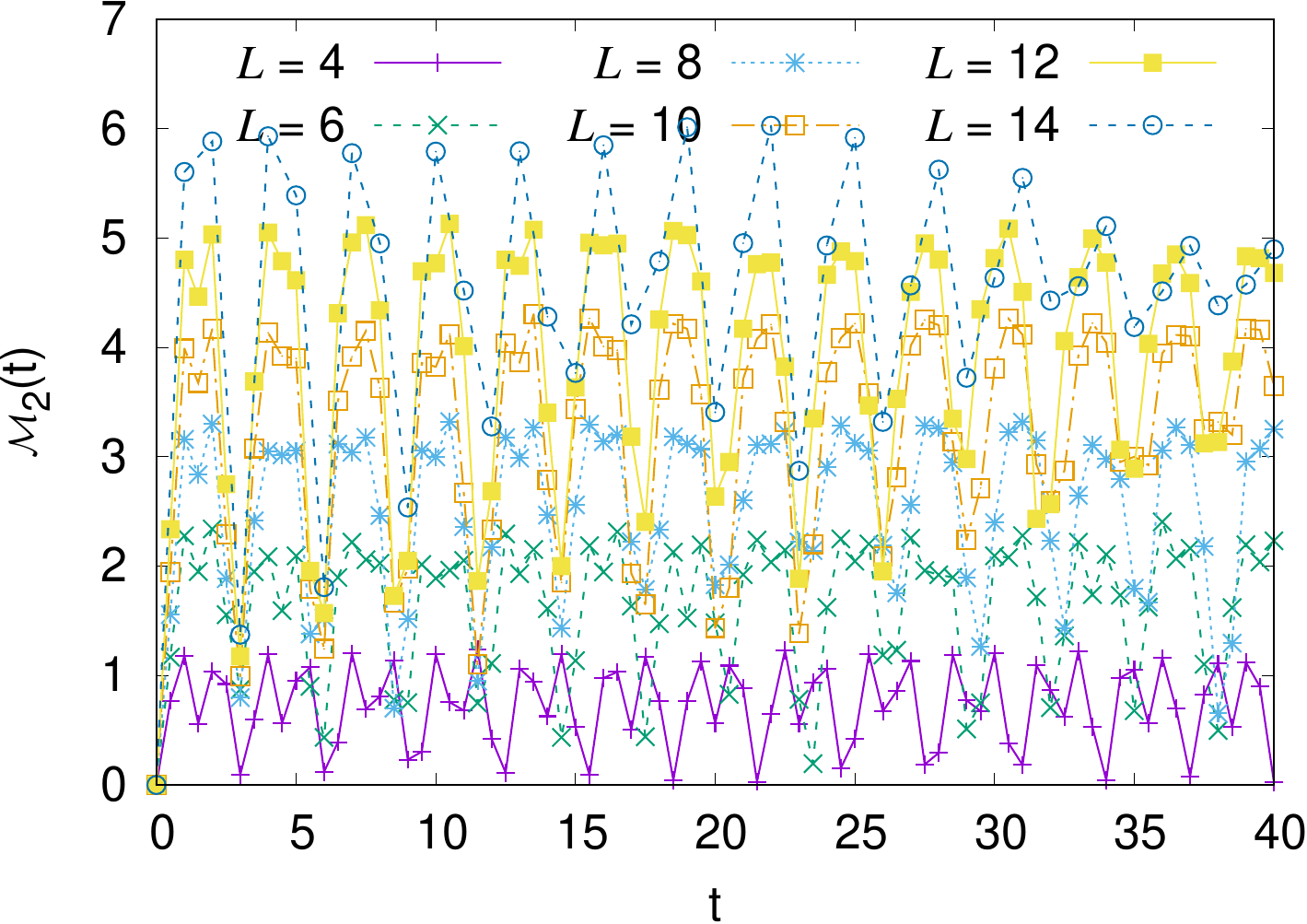}&
    \includegraphics[width=58mm]{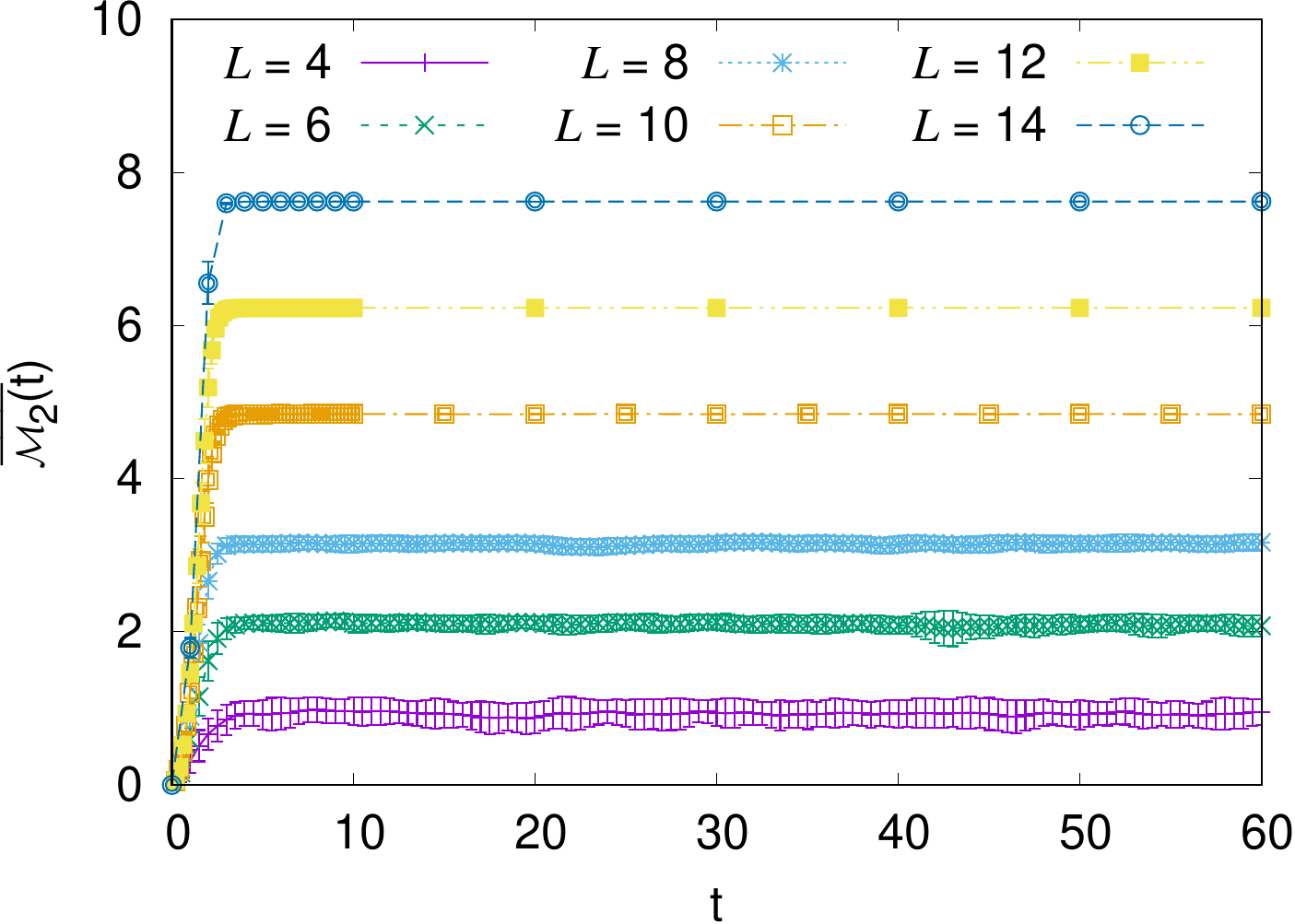}
   \end{tabular}
  \end{center}
  \caption{(a,b) The SRE [$\mathcal{M}_2(t)$ in Eq.~\eqref{orca:eqn}] vs time, for the unitary dynamics  induced by the XX-staggered Hamiltonian [Eq.~\eqref{ham:eqn} with $V=0$, $W=1$] (a), and the XXZ-staggered one [Eq.~\eqref{ham:eqn} with $V=1$, $W=1$] (b). The various curves are for different system sizes $L$ (see legend). (c) Same as (a,b) but for the SYK model in Eq.~\eqref{eq:SYK_model}, {averaged over $N_{\rm r}\geq 17$ disorder realization; the error bar is evaluated as the root-mean square deviation over realizations.}. In all panels, increasing values of SRE correspond to larger sizes $L$. We set $J=1$ in all cases, thus fixing the energy scale. To simulate the real-time dynamics, we use exact diagonalization or the Krylov technique with a time step $\delta t = 0.01$.} 
  \label{magit:fig}
\end{figure*}

We consider the dynamics of the following models.\\
{\bf (i)} The XXZ spin-1/2 chain {in a staggered field}~\cite{xing2023interactions}:
\begin{equation}\label{ham:eqn}
  \hat{H}_{\rm XXZ} \!=\!\! \sum_{j=1}^L \bigg[ \frac{J}2\left(\hat{\sigma}_j^+\hat{\sigma}_{j+1}^- \! + \! {\rm h.c.} \right) + \frac{V}{4}\hat {\sigma}_j^z\hat{\sigma}_{j+1}^z + \frac{W}{2}(-1)^j\hat{\sigma}_j^z\bigg],
\end{equation}
where $\hat{\sigma}_j^{\alpha}$ ($\alpha=x,y,z$) are the usual spin-1/2 Pauli operators, $\hat{\sigma}_j^\pm = \left(\hat{\sigma}_j^x\pm i \hat{\sigma}_j^y\right)/2$, and we assume periodic boundary conditions ($\hat \sigma^\alpha_L \equiv \hat \sigma^\alpha_1$).
In Eq.~\eqref{ham:eqn}, {the hopping term ($J$) is quadratic in the fermionic representation, the interaction term ($V$) is quartic, while the staggered field ($W$) breaks the Bethe-Ansatz integrability of the simple XXZ chain~\cite{Franchini_2017}. For $V=0$, the above Hamiltonian {boils down to a XX-staggered spin chain and} is integrable as well, since it can be mapped onto a tight-binding system of noninteracting fermions~\cite{Mbeng_SciPostPhysLectNotes24}, through a Jordan-Wigner transformation~\cite{LIEB1961407}. {The free-fermion model without staggering field was studied in the context of steady-state entanglement entropy transitions. Although the existence of a transition from area-law to logarithm-law entanglement was believed~\cite{Alberton2021,Ladewig2022}, it has been later demonstrated that the
entanglement entropy does not exhibit a phase transition in monitored free fermions, but shows a finite-size logarithmic crossover for small system sizes~\cite{Poboiko2023,Fidkowski_2021}, in agreement with the area-law behavior described by the first studies in Ref.~\cite{DeLuca2019}.}}}
\\
{\bf (ii)} The SYK model, in the spin language:
\begin{equation}
  \label{eq:SYK_model}
    \hat H_{\rm SYK}  = \frac{1}{\sqrt{L^3}} \sum_{i,j,k,l=1}^L J_{ij,kl} \, \hat{\mathcal{S}}^i\hat \sigma^+_i \, \hat{\mathcal{S}}^j\hat \sigma^+_j \, \hat{\mathcal{S}}^k\hat \sigma^-_k \, \hat{\mathcal{S}}^l\hat \sigma^-_l\,,
\end{equation}
where $\hat{\mathcal{S}}^j\equiv\prod_{\ell<j}\hat{\sigma}_\ell^z$ are the {nonlocal} string operators.
The couplings $J_{ij,kl}$ are independent Gaussian-distributed complex variables, with zero average $\langle\!\langle J_{ij,kl} \rangle\!\rangle = 0$
and variance $\langle\!\langle |J_{ij,kl}|^2 \rangle\!\rangle = J^2, \; (J \in \mathbb{R})$, satisfying $J_{ij,kl} = - J_{ji,kl} = - J_{ij,lk} = J^*_{kl,ij}$.
The $L^{-3/2}$ prefactor in front of the interaction strength guarantees that the system bandwidth is of the order of $L$, for $L \to \infty$, so that extensivity of thermodynamic quantities -- as for instance the energy -- is preserved. The SYK model in Eq.~\eqref{eq:SYK_model} is strongly quantum chaotic, being a fast scrambler of quantum information~\cite{PhysRevD.94.106002, Roberts_2018}, with all its eigenstates exhibiting volume-law entanglement entropy~\cite{Balents_2018, PhysRevD.100.041901}.

In the next sections we take these Hamiltonians and consider, from the one side, the unitary dynamics (Sec.~\ref{unidyn:sec}) and, from the other side, the monitored dynamics (Sec.~\ref{monidyn:sec}) performed with a quantum-trajectory approach (so that the state of the system is always pure). In both cases, we evaluate the nonstabilizerness on the evolving quantum state $\ket{\psi_t}$ at time $t$, defined through the SRE~\cite{leone2022, oliviero2022}
\begin{equation}\label{orca:eqn}
  \mathcal{M}_2(t) = -\ln \bigg[ \frac{1}{2^L}\sum_{\hat{P}\in\mathcal{P}_L}\braket{\psi_t|\hat{P}|\psi_t}^4\bigg]\,,
\end{equation}
where we consider the set $\mathcal{P}_L$ of the Pauli strings
%
$\hat{P}=\prod_{j=1}^L\hat{\sigma}_j^{\alpha_j}$,
%
with $\alpha_j=0,\,1,\,2,\,3$ ($\hat{\sigma}_j^0 \equiv \hat{\mathbb{I}}$, $\hat{\sigma}_j^1 \equiv \hat{\sigma}_j^x$, $\hat{\sigma}_j^2 \equiv \hat{\sigma}_j^y$, $\hat{\sigma}_j^3 \equiv \hat{\sigma}_j^z$). From a physical point of view, the SRE is a delocalization measure of the state seen as an operator in the basis of the Pauli operators, and has a structure similar to the inverse participation ratio that evaluates the delocalization of a state in a given basis of states~\cite{Edwards_JPC72}. In Appendix~\ref{nonstabilizerness:sec} we provide details on the numerical evaluation of the SRE.
%

\section{Unitary dynamics}\label{unidyn:sec}

Let us start with the unitary dynamics.
For both Hamiltonians~\eqref{ham:eqn} and~\eqref{eq:SYK_model}, we initialize the system in the factorized N\'eel state $\ket{\up,\down,\up,\down,\ldots,\up,\down}$ and {perform the unitary dynamics using} exact diagonalization or the Krylov technique as implemented in Expokit~\cite{Sidje_Expokit}. 
The unitary dynamics of both models conserves the total $z$ magnetization $\hat{S}^z=\frac12\sum_j\hat{\sigma}_j^z$, and we restrict therefore our dynamics to the invariant subspace with $\hat{S}^z \equiv 0$, whose dimension is $\mathcal{N}_L=\binom{L}{L/2}$.

Some examples of the time evolution of the SRE~\eqref{orca:eqn} are shown in Fig.~\ref{magit:fig}: panels (a) and (b) are for the XXZ-staggered chain with $V=0$ and $V=1$ respectively, while panel (c) is for the SYK model.
We see that, in the SYK case, there are very small fluctuations around an average value, in stark contrast with the wide oscillations of the XXZ-staggered case. {The results for the SYK model are averaged over $N_{\rm r}$ disorder realizations and the error bars are evaluated as the root-mean square deviation. For $L\geq 10$ the error bars become negligible, marking that all the realizations behave essentially in the same way.} The SYK model is strongly quantum chaotic and displays a self-averaging behavior that erases fluctuations {and the dependence on the disorder realization}, which is also seen in the behavior of the SRE. 


\begin{figure}[t]
  \begin{center}
   \begin{tabular}{c}
    (a) \\
    \includegraphics[width=80mm]{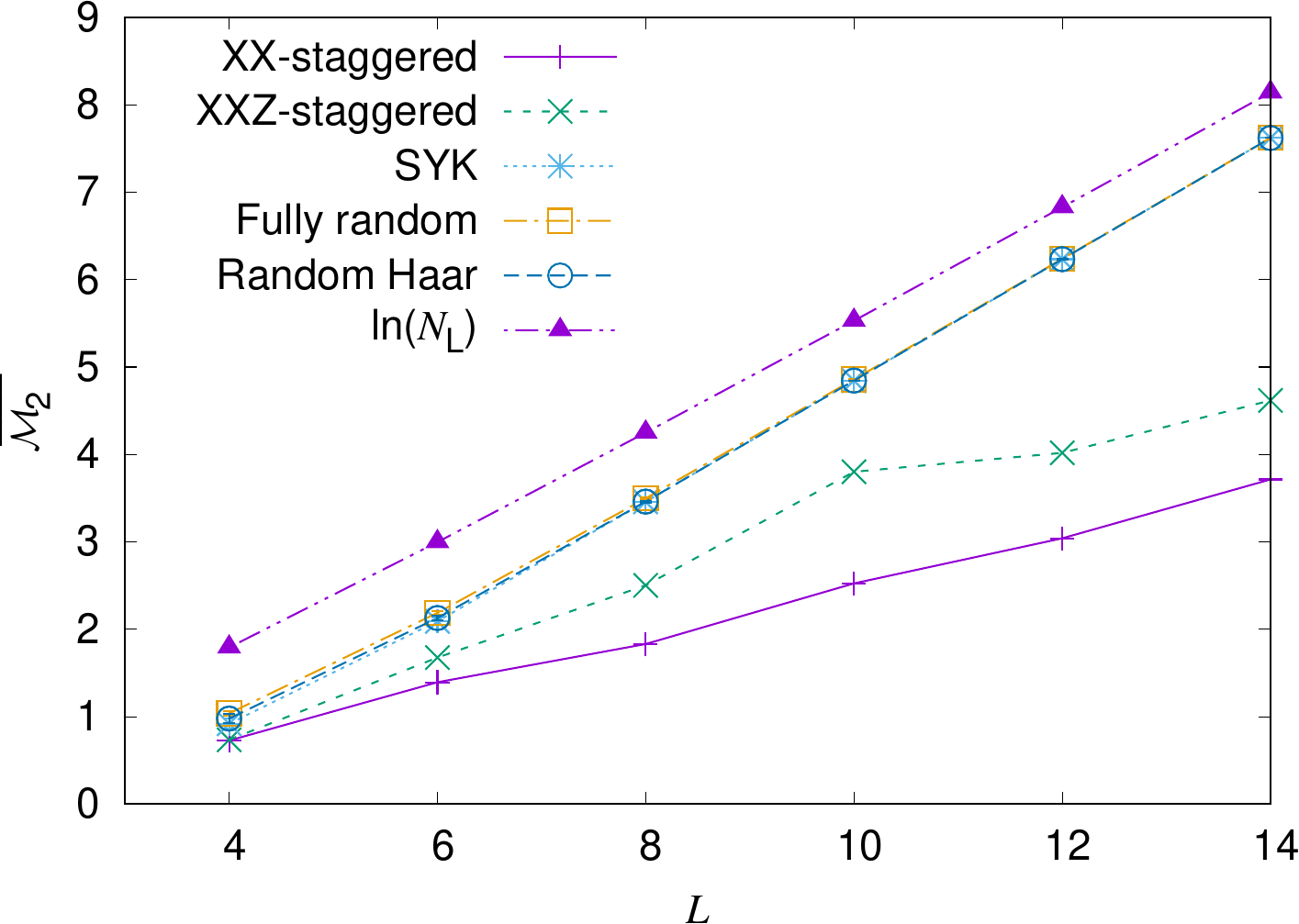}\\
    (b) \\
   \includegraphics[width=80mm]{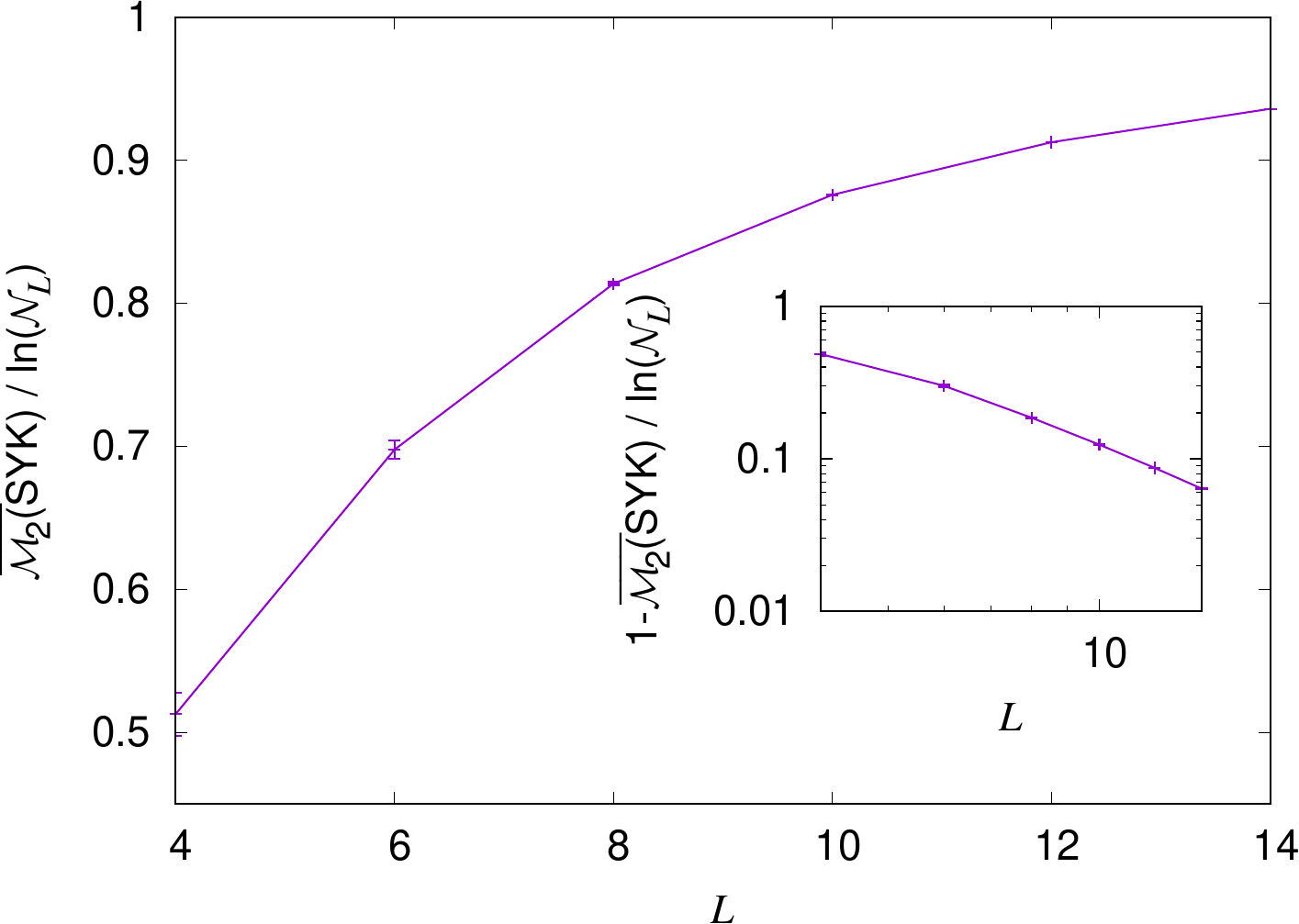}
   \end{tabular}
  \end{center}
  \caption{(a) The time-averaged SRE $\overline{\mathcal{M}_2}$ vs $L$, for the XX-staggered, the XXZ-staggered, and the SYK model, compared with the corresponding results for the {random-phase} {and Haar} random states. For reference, we also plot the {theoretically predicted~\cite{leone2021} quantum-chaotic behavior} $\overline{\mathcal{M}_2} \sim \ln (\mathcal{N}_L)$. Data for the SYK model and for the {random-phase} {and Haar} random states are averaged over $N_{\rm r}\geq 11$ disorder realizations; {here the error bars on the average (not visible on this scale) are evaluated as the root mean square deviation divided by $\sqrt{N_{\rm r}}$}. (b) The ratio between $\overline{\mathcal{M}_2}$ in the SYK model and the theoretically predicted SRE quantum chaotic value $\ln (\mathcal{N}_L)$ vs $L$. (Inset) 1 minus the ratio versus $L$ in double-logarithmic scale.}\label{ratio:fig}
\end{figure}

Looking at the data in Fig.~\ref{ratio:fig}(a),
which reports the time average of the SRE versus the system size for the different models, we can observe that in all cases this time average increases approximately linearly with $L$.
{More specifically, it is larger in the nonintegrable XXZ-staggered case ($V=1$) rather than in the integrable free-fermion limit corresponding to a XX-staggered ($V=0$) chain, while it attains its maximum value for the SYK system.} 

We compare these curves with the SRE evaluated for random states, as anticipated above. {We begin with the {random-phase state},} given by
%
  $\ket{\psi} = \frac{1}{\sqrt{\mathcal{N}_L}}\sum_{\{s_j\}}\nep^{-i\varphi_{\{s_j\}}}\ket{\{s_j\}}$,
%
where the sum is performed over the configurations such that $\sum_{j=1}^Ls_j = 0$, and the $\varphi_{\{s_j\}}$ are random numbers uniformly distributed in $[0,2\pi)$. Data are averaged over different realizations of the random state. {We also consider the average of the SRE over random Haar states. These states are constructed as the columns of random matrices distributed according the Haar measure, that is invariant under all unitary transformations~\cite{Mezzadri}. The difference {with} the {random-phase state} value turns out to be negligible and decreases for increasing system sizes. Notice that this value is smaller than that for random Haar states~\cite{e26060471,PhysRevB.111.054301}, because here we are restricting to the Hilbert subspace with a vanishing total $z$ magnetization.}

We see from Fig.~\ref{ratio:fig}(a) that the time-averaged SRE for the SYK model saturates the bound provided by the random states, when $L \gtrsim 10$. It is {not strange }that this result is recovered at such small system sizes, as the SYK model displays strongest quantum chaos. {Furthermore}, both the averaged SRE of the SYK model and the one of the random states tend to be asymptotically parallel to the value $\ln\mathcal{N}_L$ given by the prediction of Ref.~\cite{leone2021} for quantum chaotic systems. {Figure~\ref{ratio:fig}(b) reports the ratio between the time-averaged SRE of the SYK model and the theoretical quantum-chaotic prediction $\ln\mathcal{N}_L$, which seems to approach the unit value for increasing $L$.} 

\section{Monitored dynamics}\label{monidyn:sec}
\begin{figure*}[t]
  \begin{center}
   \begin{tabular}{ccc}
       (a) XX-staggered&(b) XXZ-staggered&(c) SYK\\
       \includegraphics[width=58mm]{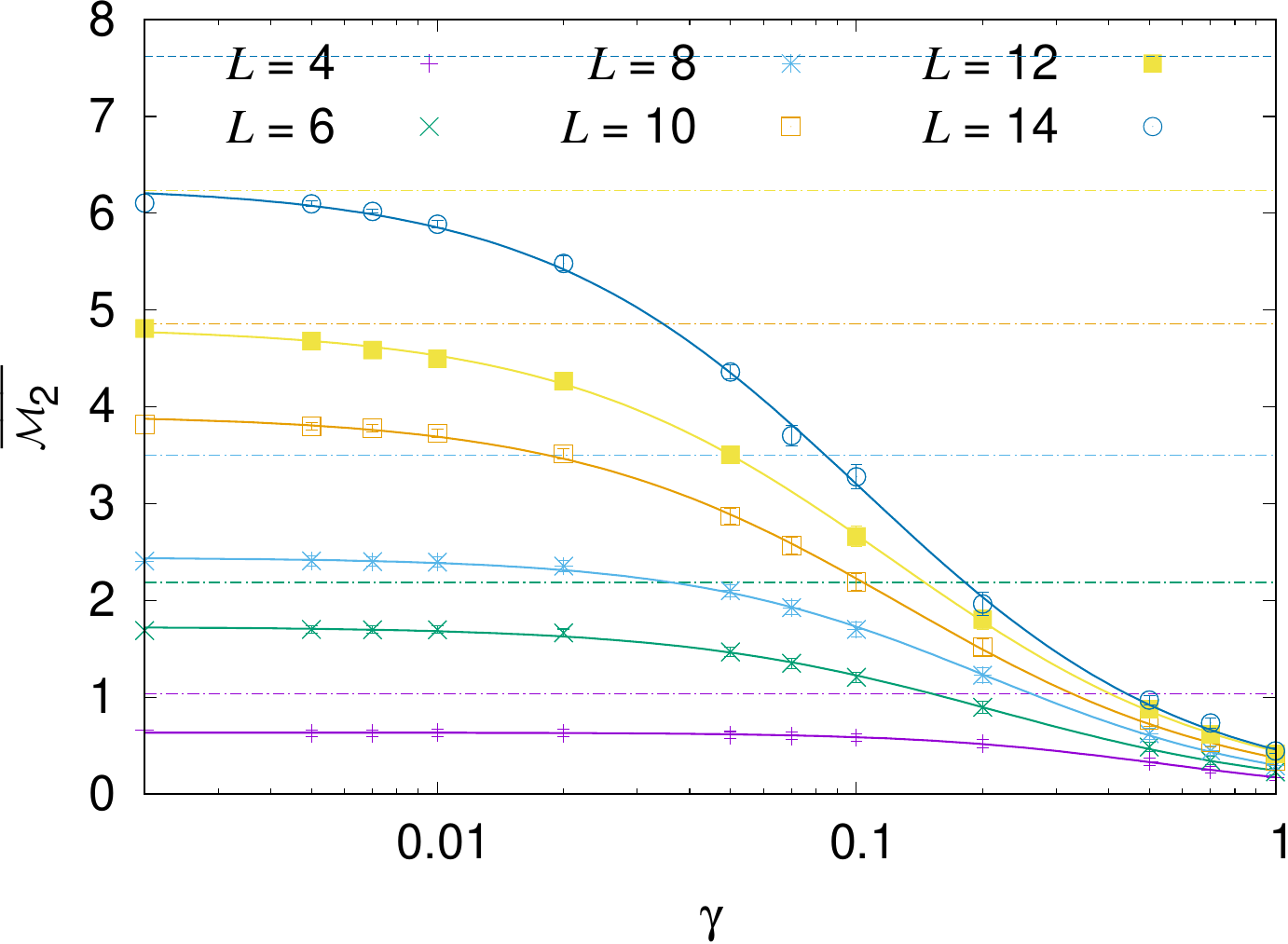}&
       \includegraphics[width=58mm]{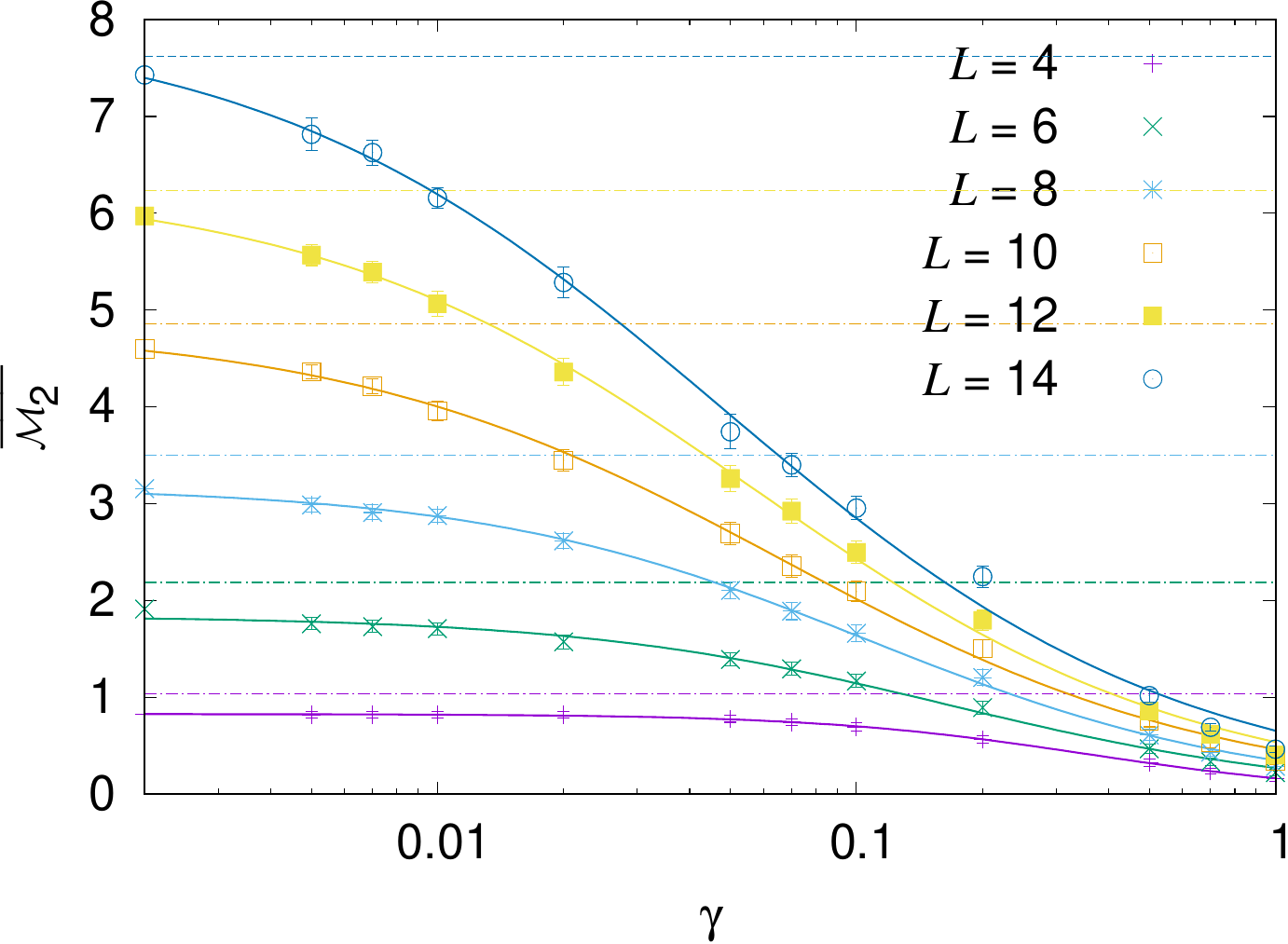}&
       \includegraphics[width=58mm]{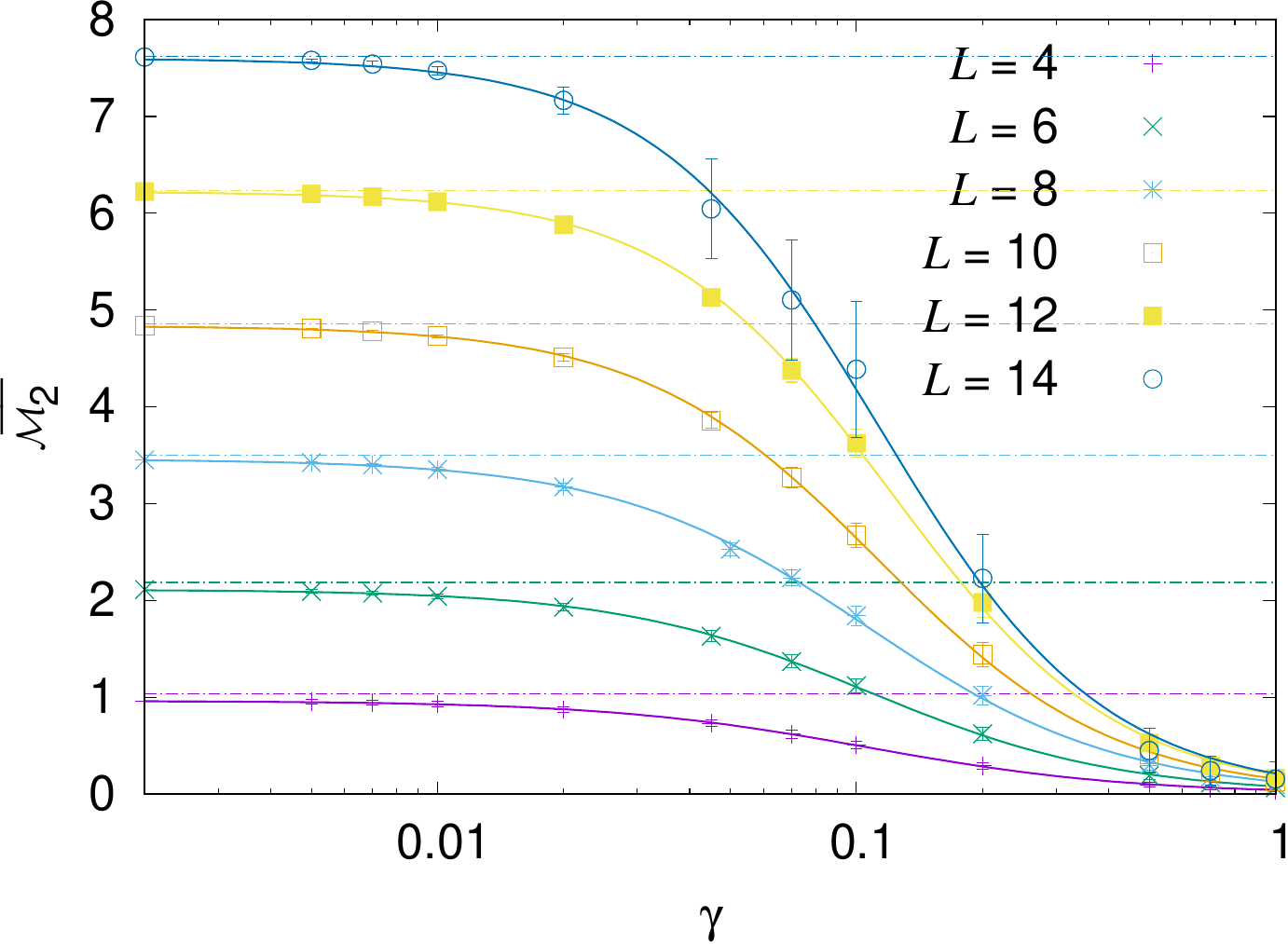}
   \end{tabular}
  \end{center}
  \caption{The monitored-dynamics {steady-state} SRE $\overline{\mathcal{M}_2}$, obtained by averaging over time and trajectories, vs the coupling with the environment $\gamma$, for the three models of Fig.~\ref{magit:fig} with the same parameters. 
  The various data sets are for different values of $L$ (see legend).
  In all panels, increasing values of the SRE correspond to larger $L$.
  Horizontal dashed lines indicate the {random-phase state} value for the corresponding $L$. The fits to the numerical data (symbols), obtained with the least-square method applied to the function in Eq.~\eqref{fl:eqn}, are reported with continuous lines.  
  The color code for the various quantities follows the corresponding values of $L$.
  We fix the energy scale by setting $J = 1$ in all models. 
  Data for $L \leq 12$ are averaged over $N_{\rm r}=48$ trajectories, while for $L=14$ we average over $N_{\rm r}\geq 11$ realizations.
  }
  \label{mitgam:fig}
\end{figure*}
We now move to study the SRE in the monitored dynamics. We apply to both models a QSD measurement protocol as in Ref.~\cite{DeLuca2019}. Here we choose to probe the onsite $z$-magnetization operators $\hat{\sigma}_j^z$ as in Ref.~\cite{xing2023interactions}. 
{Discretizing the time in steps of $\delta t$, we get the stochastic-Hamiltonian evolution
\begin{eqnarray}
  \ket{\psi_{t+\delta t}} & = & \ket{\psi_t} - i \hat{H}\delta t\ket{\psi_t} + \sum_{l=1}^L\Big[\delta\xi_l(t)\left(\hat{\sigma}_l^z-\braket{\sigma_l^z}_t\right)\nonumber\\
    &&-\frac{\gamma}{2}\delta t\left(\hat{\sigma}_l^z-\braket{\sigma_l^z}_t\right)^2\Big]\ket{\psi_t}+ \mathcal{O}(\gamma^2\delta t^2)\,,
    \label{co:eqn}
\end{eqnarray}
where $\delta\xi_l(t)$ are uncorrelated random Gaussian variables with vanishing average and variance $\gamma\delta t$, and we have defined $\braket{\sigma_l^z}_t \equiv \braket{\psi_t|\hat{\sigma}_l^z|\psi_t}$. The unitary dynamics of both models~\eqref{ham:eqn} and~\eqref{eq:SYK_model} conserves the total $z$ magnetization $\hat{S}^z=\frac12\sum_j\hat{\sigma}_j^z$, thus we restrict to the subspace $\mathcal{N}_L$ with $\hat{S}^z = 0$. Adding the {stochastic noise due to quantum measurements}, this conservation law is kept unchanged, and Eq.~\eqref{co:eqn} can be Trotterized as~\cite{DeLuca2019}.
\begin{equation} \label{psicola:eqn}
  \ket{\psi_{t+\delta t}} = \mathcal{N}\nep^{\sum_l\hat{\sigma}_l^z\left[\delta\xi_l(t)+2 \gamma \delta t \, \braket{\sigma_l^z} \right]}\nep^{-i\hat{H}\delta t}\ket{\psi_t} ,
\end{equation}
where $\mathcal{N}$ is a normalization factor~\cite{notap}.}

{This quantum-trajectory approach describes the so-called QSD protocol and was introduced in Ref.~\cite{Gisin_1992}. It can be experimentally realized in quantum optical systems by coupling the system subsequently to a series of identically prepared ancillas and repeating a POVM measure with the same {Kraus operators}~\cite{gardi:book} (see also the discussion in Ref.~\cite{Turkeshi2021}). Ref.~\cite{Plenio} explains the connection with homodyne measurements in a framework where the difference of the photon count between two photodetectors is measured. The detectors act on the two branches of a beamsplitter that mixes radiation from a cavity in a generic quantum state and a cavity in a coherent state.} 

{Averaging over the trajectories, and performing the limit $\delta t\to 0$, one gets the Lindblad equation for the average density matrix $\rho(t) = \overline{\ket{\psi_t}\bra{\psi_t}}$, 
\begin{equation}
  \partial_t\rho_t \!= -i [\hat{H}, \rho_t]
  + \gamma \!\sum_j \!\Big( \hat{\sigma}_j^z \, \rho_t \, \hat{\sigma}_j^z -\rho_t \Big)\,.
  \label{eq:lind}
\end{equation}
In this sense, the QSD protocol in Eq.~\eqref{co:eqn} is an unraveling of this Lindblad equation. Evaluating linear observables along the trajectories and then averaging provides the same result as with the Lindblad equation, due to the linearity of the observables on the quantum state. The situation is different for the SRE, as it is nonlinear in the quantum state $\ket{\psi_t}$.}

Each realization of the stochastic evolution provides a single quantum trajectory, along which we evaluate the SRE, $\mathcal{M}_2(t)$ in Eq.~\eqref{orca:eqn}. We mark with $\overline{\mathcal{M}_2}(t)$ the average of the SRE over $N_{\rm r}$ quantum trajectories at time $t$. Then we average over time, between an initial and a final value chosen such that {the steady state} has been attained. 
The SRE averaged over realizations and time
is denoted by $\overline{\mathcal{M}_2}$ and quantifies the {steady-state} value of $\overline{\mathcal{M}_2}(t)$. The error bars on the average are computed as the root mean square deviation divided by $\sqrt{N_{\rm r}}$, the square root of the number of realizations. We still initialize the dynamics in the N\'eel state $\ket{\up,\down,\up,\down,\ldots,\up,\down}$ and use the same numerical techniques adopted to study the unitary dynamics. We remark that, for the XX-staggered chain ($V=0$), even the monitored dynamics is integrable and can be shown to be equivalent to a free-fermion case~\cite{DeLuca2019}.

Our numerical results for $\overline{\mathcal{M}_2}$ versus $\gamma$ are reported in Fig.~\ref{mitgam:fig}. Panel (a) is for the integrable XX-staggered chain [Eq.~\eqref{ham:eqn} with $V=0$].
We see that, for each value of $L$, the curve lies much below the value corresponding to the {random-phase state} (horizontal dashed line with the same color code). On the other hand, the curves reported in Fig.~\ref{mitgam:fig}(b) for the nonintegrable XXZ-staggered chain [Eq.~\eqref{ham:eqn} with $V=1$] tend to the corresponding random-state value for $\gamma\to 0$. 
The same occurs for the SYK model of Eq.~\eqref{eq:SYK_model}, where convergence is attained for a larger value of $\gamma$ [Fig.~\ref{mitgam:fig}(c)].

At this stage, we emphasize that there is an important difference between the two cases (b) and (c): in fact, while in the SYK model the $\gamma\to 0$ limit coincides with the averaged SRE in the unitary case shown in Fig.~\ref{ratio:fig}(a), in the nonintegrable XXZ-staggered chain it does not. In this latter case, only the limit $\gamma\to 0$ of the {steady-state} SRE coincides with the result for the {random-phase state}, while the time average in the unitary case does not. Therefore the $\gamma\to 0$ limit is singular.
{The reason is that, in the interacting XXZ-staggered chain, the presence of a small amount of {stochastic noise due to quantum measurements} disrupts the conservation of the energy (i.e., the only {local} integral of motion of this system in the subspace of interest), thus allowing the relaxation to the random-state value.}

{In contrast with the spin-chain model}, the Hamiltonian of the SYK model is highly nonlocal, and its conservation does not provide a strong constraint to the dynamics. Breaking this conservation with an infinitesimal {stochastic noise due to quantum measurements}, therefore, {has a negligible effect on the steady-state behavior of the SRE}: The $\gamma\to 0$ limit coincides with the averaged SRE in the unitary case, and both coincide with the random-state value.

\begin{figure*}[t]
  \begin{center}
   \begin{tabular}{ccc}
       (a) $A_L$&(b) $b_L$&(c) $\gamma_{0,L}$\\
       \includegraphics[width=58mm]{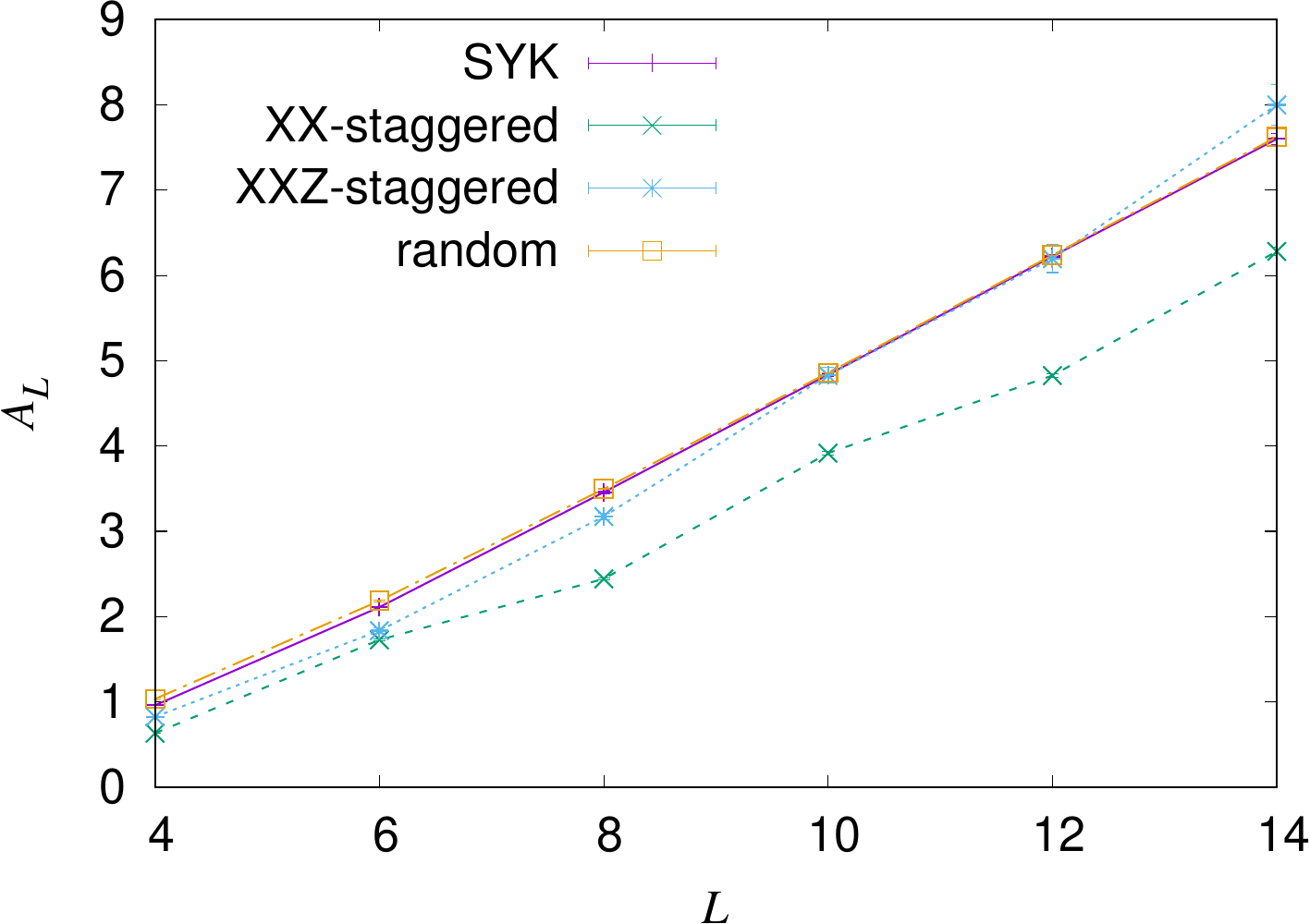}&
       \includegraphics[width=58mm]{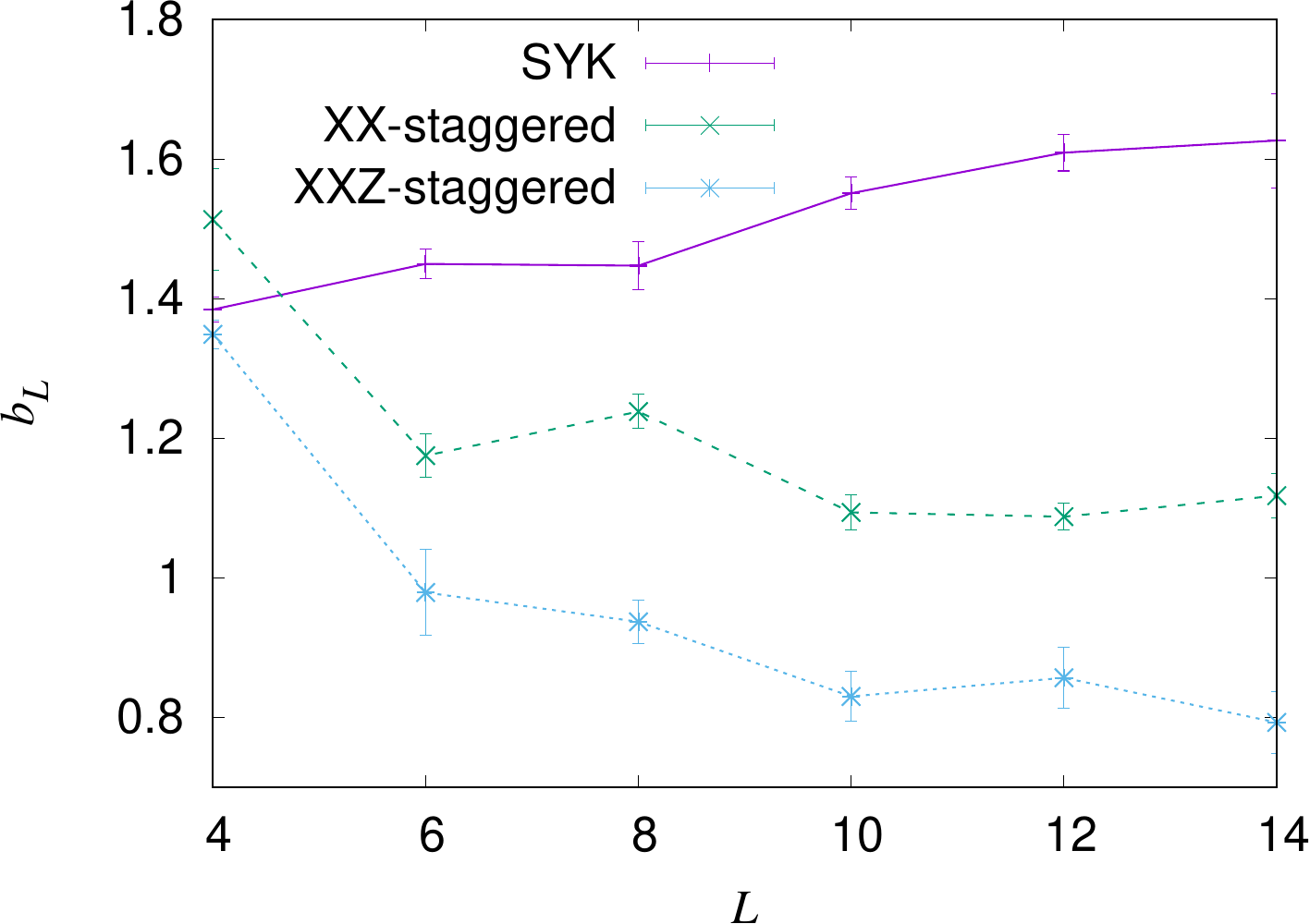}&
       \includegraphics[width=58mm]{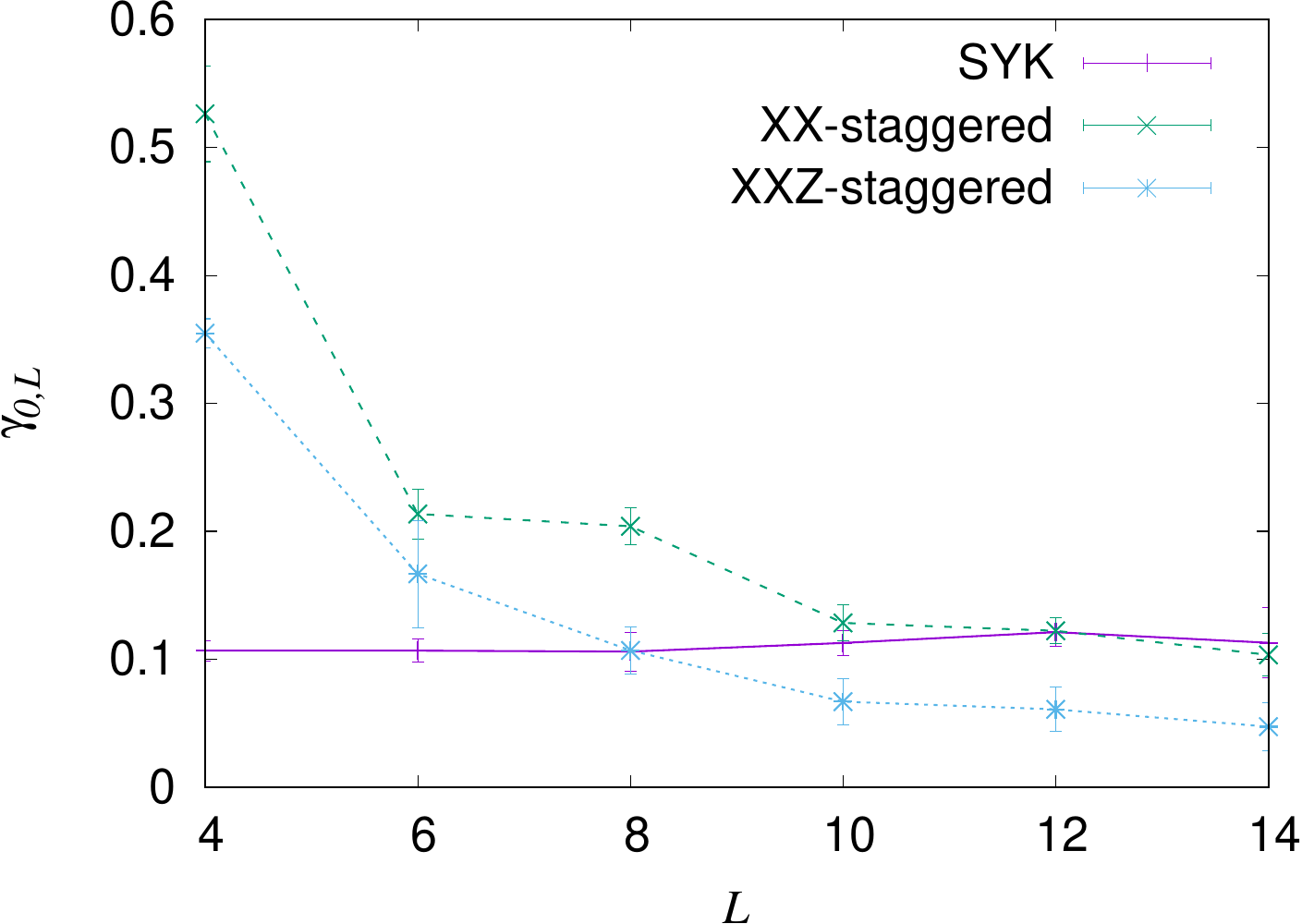}
   \end{tabular}
  \end{center}
  \caption{Parameters of the fit with Eq.~\eqref{fl:eqn}, for the {steady-state} SRE $\overline{\mathcal{M}_2}$ vs $\gamma$, plotted against $L$. 
  Specifically we plot $A_L$ (a), $b_{L}$ (b), and $\gamma_{0,L}$ (c).
  The various data sets are for different models, as indicated in the legend. In (a) we also plot, for comparison, the {random-phase state} value of the {SRE}. Notice the linear increase in (a) and  the saturation towards an asymptotic value in (b) and (c). 
  }\label{paraf:fig}
\end{figure*}

\begin{figure}[th!]
  \begin{center}
   \begin{tabular}{ccc}
       (a) XX-staggered\\
       \includegraphics[width=80mm]{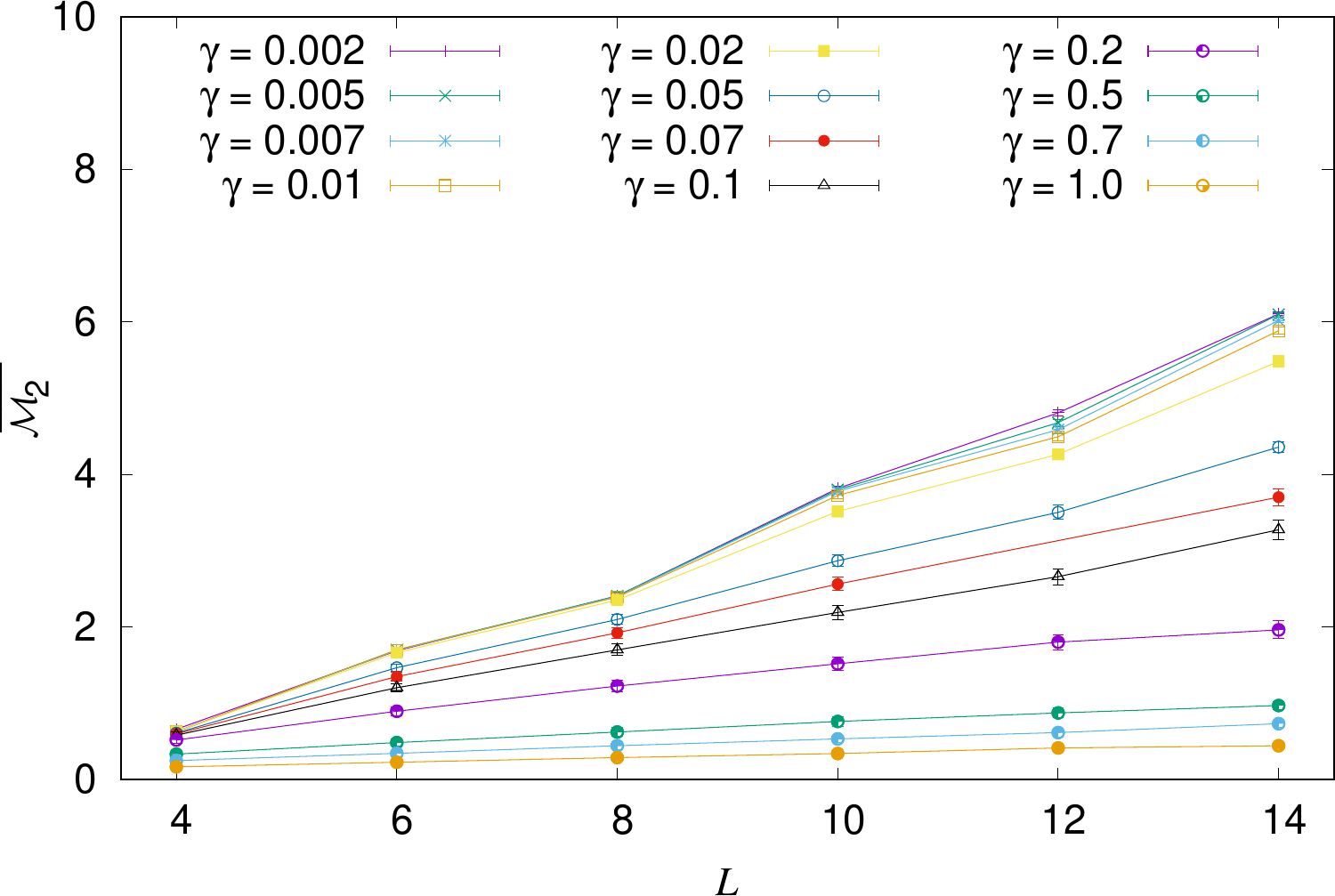}\\
       (b) XXZ-staggered\\
       \includegraphics[width=80mm]{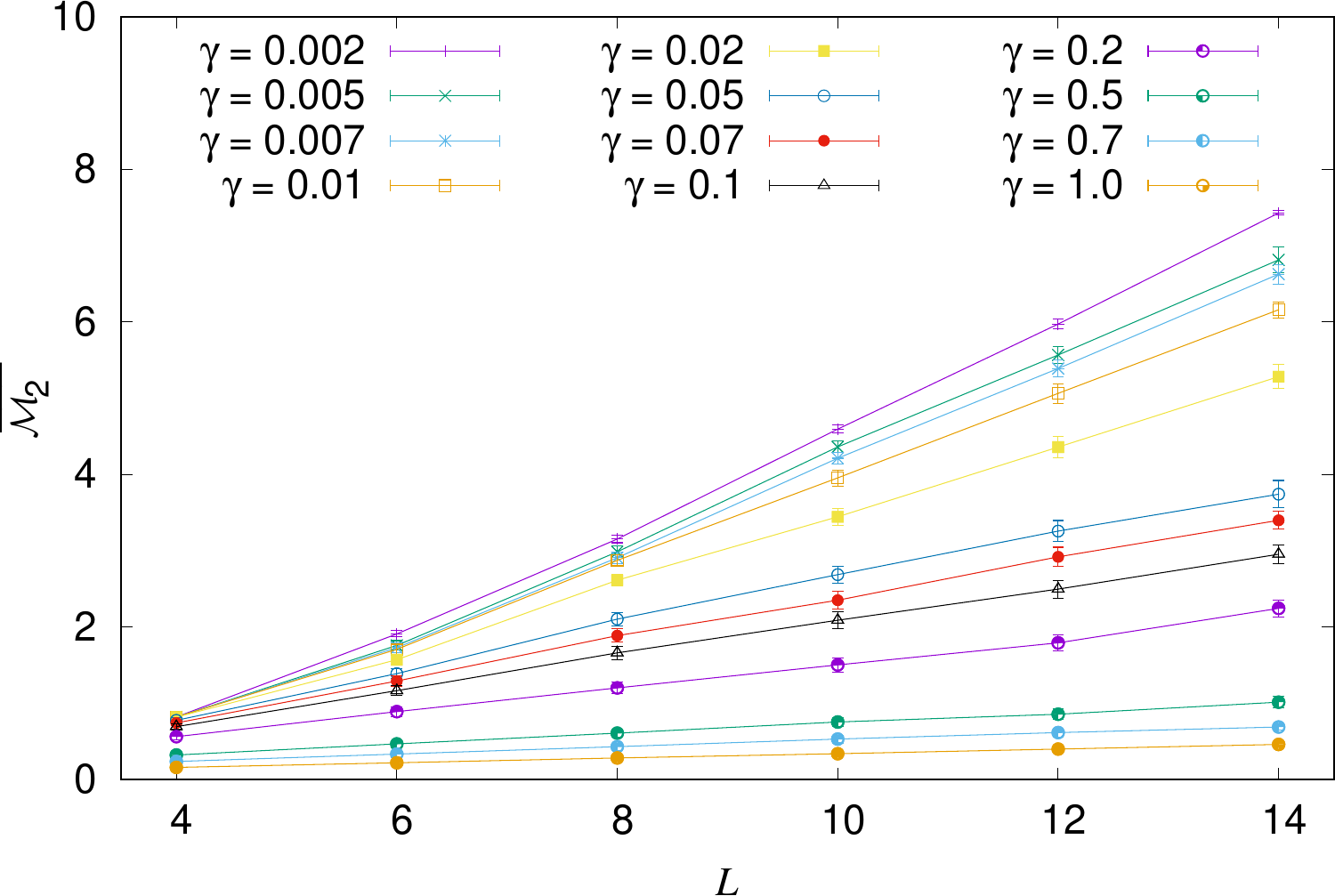}\\
       (c) SYK\\
       \includegraphics[width=80mm]{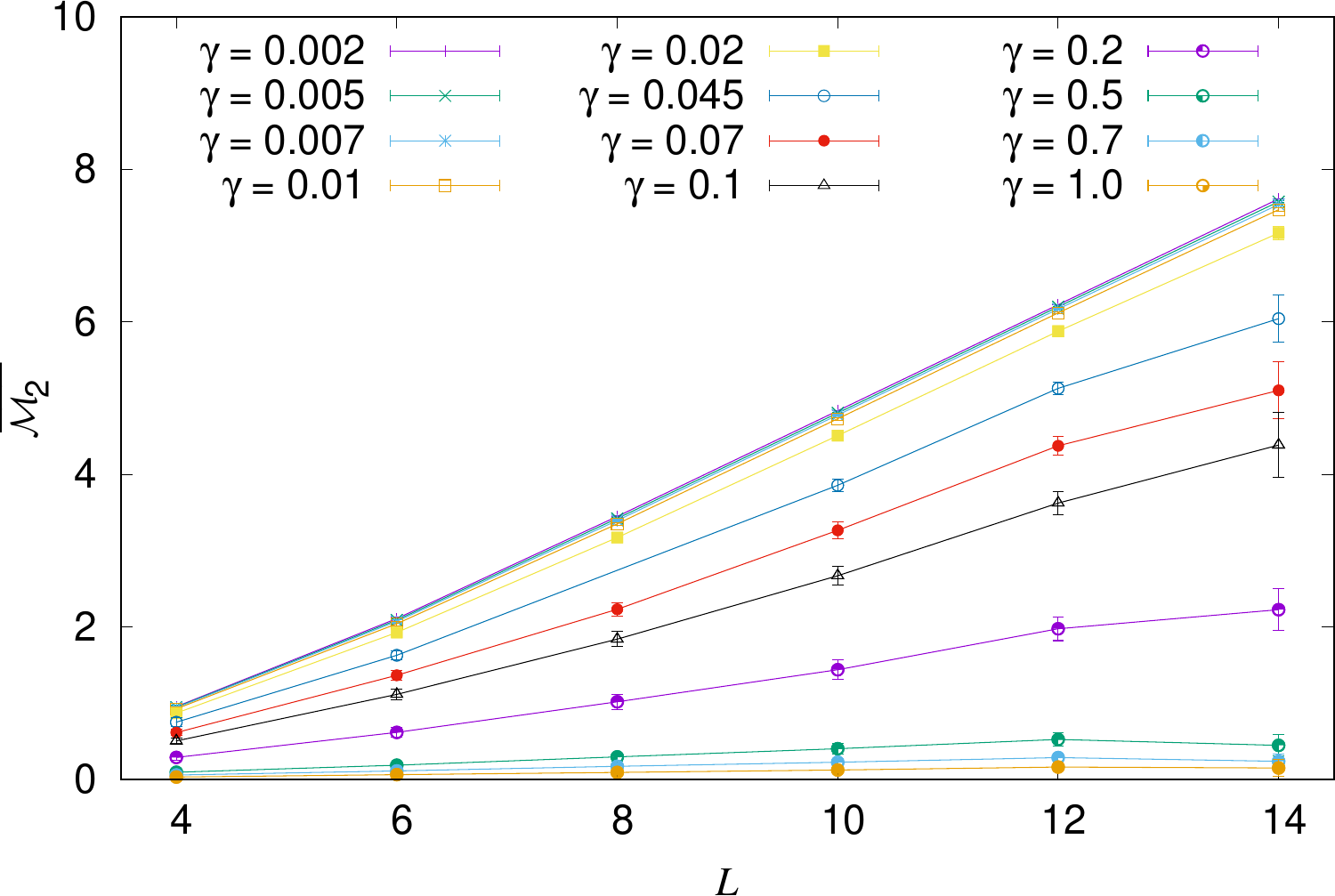}
   \end{tabular}
  \end{center}
  \caption{The {steady-state} SRE $\overline{\mathcal{M}_2}$ in the monitored-dynamics, obtained by averaging over time and trajectories, vs $L$, for the three models of Fig.~\ref{mitgam:fig} with the same parameters. Panel (a) refers to the XX-staggered model, panel (b) to the XXZ-staggered model, and panel (c) to the SYK model.
  The various data sets are for different values of $\gamma$ (see legend).
  Data for $L \leq 12$ are averaged over $N_{\rm r}=48$ trajectories, while for $L=14$ we average over $N_{\rm r}\geq 11$ realizations.
  }
  \label{mitL:fig}
\end{figure}

In all the considered cases, we can fit the curves of $\overline{\mathcal{M}_2}$ versus $\gamma$ with the {generalized Lorentzian} function 
\begin{equation}\label{fl:eqn}
  f_L(\gamma) = \frac{A_L}{1 + ( \gamma / \gamma_{0,L} )^{b_L}} \,,
\end{equation}
which reduces to the usual Lorentzian for $b_L = 2$. {In this formula $A_L$ denotes the behavior of the magic for $\gamma\to 0$, $\gamma_{0,L}$ the range over which it decays, and $b_L$ provides the steepness with which this happens. }We report the result of the fitting procedure in the various panels of Fig.~\ref{mitgam:fig} by the continuous curves, which reliably fit all the symbols denoting numerical points. We {could not identify the physical origin of why the fitting function reproduces the data so well}, and this will be the focus of further studies.

The values of the fitting parameters $A_L$, $b_L$, and $\gamma_{0,L}$ versus the size $L$ are reported in Fig.~\ref{paraf:fig}. We {can} see that $A_L$ is linear in $L$ and, for the nonintegrable cases, closely follows the {random-phase state} value of the {SRE} [Fig.~\ref{paraf:fig}(a)]. This is meaningful because $A_L$ is the limit for $\gamma\to 0$ of $\overline{\mathcal{M}_2}$. {While in the interacting XXZ-staggered chain ($V \neq 0$) a small {stochastic noise due to quantum measurements} restores the relaxation to the random-state value, as discussed above, this cannot happen for the noninteracting XX-staggered chain ($V=0$), due to its equivalence with an integrable system of quadratic fermions (studied in Ref.~\cite{DeLuca2019}), and indeed in this case $A_L$ does not saturate the random-state value.}
Moreover, we see that the behaviors of $b_L$ [Fig.~\ref{paraf:fig}(b)] and of $\gamma_{0,L}$ [Fig.~\ref{paraf:fig}(c)] are consistent with an asymptotic convergence to a finite value, for increasing $L$. {In the SYK case, the exponent $b_L$ converges to a larger value, taking into account the steeper descent of the curves in Fig.~\ref{mitgam:fig}(c). In that panel we also notice that fluctuations around the average value are maximum for $\gamma \approx 0.1$ and decrease moving towards the unitary ($\gamma\to 0$) or the factorized ($\gamma\to\infty$) limit.}

Summarizing, {since $A_L$ is linear in $L$, while $b_L$ and $\gamma_{0,L}$ are asymptotically constant in $L$}, our data suggest that, in all the considered models, the {{steady-state}} SRE tends to a linear increase with the system size [embodied by $A_L$ in Eq.~\eqref{fl:eqn}], whatever the value of the coupling with the environment. {For large $\gamma$, the SRE tends to 0 as a power law with exponent $-b_L$, and the quantum-trajectory ensemble is drained of any quantum complexity. }

{In order to corroborate these results, in Fig.~\ref{mitL:fig} we plot $\overline{\mathcal{M}_2}$ vs $L$ for different values of $\gamma$ in the three models. In the XXZ-staggered [Fig.~\ref{mitL:fig}(b)] and the SYK [Fig.~\ref{mitL:fig}(c)] models, the behavior is always consistent with a linear increase, within the errobars. In the SYK model the curves tend to bend downwards for $L=14$ and $\gamma\geq 0.045$, but this could be an artifact due to the smaller available sampling for such large size. In the XX-staggered model [Fig.~\ref{mitL:fig}(a)] the situation is more complicated, because for $\gamma<0.05$ there are some oscillations superimposed to the linear increase. This finding is consistent with the results of Ref.~\cite{tirrito2025magicphasetransitionsmonitored}, where, in the XX model with $W=0$, logarithmic corrections to the linear increase are found that disappear for $\gamma$ above a threshold. It might also be related to a transition in the magic mutual information~\cite{wang2025magictransitionmonitoredfree}. We fit the curves of $\overline{\mathcal{M}_2}$ vs $L$ in Fig.~\ref{mitL:fig} with straight lines, take the slope $m$ resulting from the linear minimum-square fit, and plot it versus $\gamma$ in Fig.~\ref{slope:fig} for all the considered models. Quite remarkably, the curves $m$ vs $\gamma$ for the SYK and the XX-staggered models can also be fitted with a generalized Lorentzian function as given in Eq.~\eqref{fl:eqn}, while the same does not happen for the XXZ-staggered model. In both the XXZ-staggered and the SYK models the slope $m$ attains the value $\ln 2$ -- the one corresponding to the random-state case~\cite{e26060471} -- in the limit $\gamma\to 0$.
\begin{figure}[t]
  \begin{center}
   \begin{tabular}{c}
    \includegraphics[width=80mm]{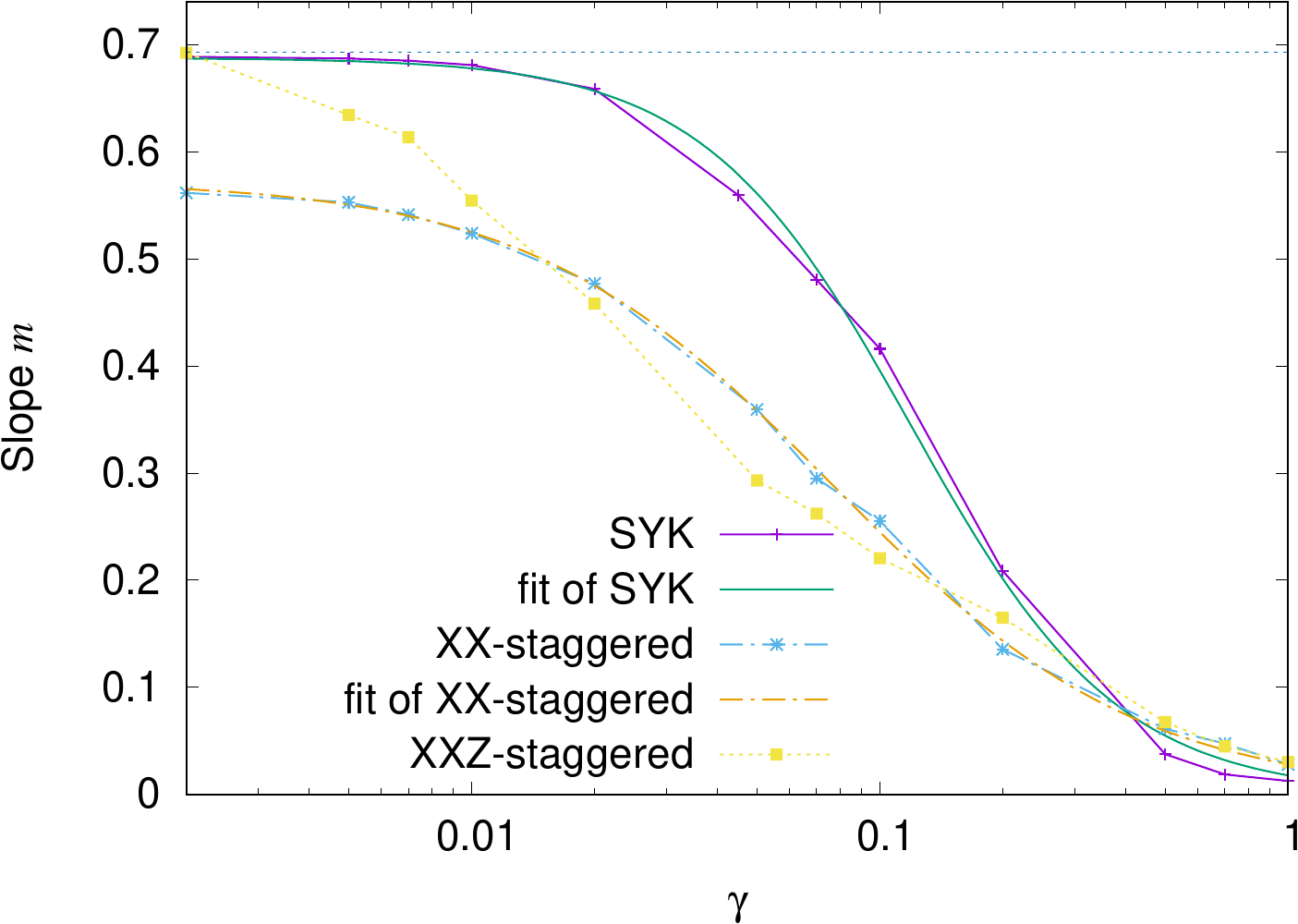}\\
   \end{tabular}
  \end{center}
  \caption{Slope $m$ of $\overline{\mathcal{M}_2}$ versus $L$ obtained with the linear minimum-square fit of the curves in Fig.~\ref{mitL:fig} and plotted vs $\gamma$ with a logarithmic scale on the horizontal axis. The SYK curve and the XX-staggered curve can be reasonably fitted with the generalized Lorentzian Eq.~\eqref{fl:eqn}. The horizontal line marks the value $\ln 2$ that is attained by the XXZ-staggered and the SYK curves in the limit $\gamma\to 0$, and coincides with the random-state value.}\label{slope:fig}
\end{figure}
}
\section{Conclusion}\label{conc:sec}

We have considered the monitored dynamics of many-body quantum systems undergoing continuous measurements, under a QSD unraveling. We have focused on two Hamiltonians, the XXZ spin chain in a staggered field and the SYK model, where the former can be made interacting or free-fermion integrable by changing a single parameter [either $V$ or $W$ in Eq.~\eqref{ham:eqn}], and the latter in Eq.~\eqref{eq:SYK_model} describes the most chaotic quantum system. We have evaluated the SRE along each quantum trajectory, as a probe of quantum-complexity properties different than entanglement.

We have first evaluated the SRE in the case of a pure unitary evolution. We have found that only in the SYK model---that is a fast scrambler and a strongly chaotic quantum system---the SRE after {an initial transient behavior} attains a {steady-state} value, and fluctuations in time around this value are negligible. Considering the time average, we see that in all models there is a behavior consistent with a linear increase in the system size. Only the SYK model saturates the bound provided by the ({random-phase} or Haar) random state when $L\geq 8$, and the SRE in this case tends to become asymptotically parallel to the logarithm of the Hilbert space dimension, as predicted for chaotic systems~\cite{leone2021}.

Then we have moved to the monitored case. Here the SRE averaged over the quantum trajectories reaches a {{steady-state}} value after {an initial transient behavior}. For both models, we have studied this {{steady-state}} value for different system sizes and found that its dependence on $\gamma$, the coupling to the measuring environment, can be effectively fitted by a {generalized Lorentzian} function. {In the limit of small $\gamma$, the {{steady-state}} SRE in the nonintegrable XXZ-staggered chain recovers the random-state value. In this case the limit is singular, because the {stochastic noise due to quantum measurements} for small but nonvanishing $\gamma$ breaks the conservation of the energy, the only local integral of motion of this system in the subspace of interest. Conversely, the integrable (free-fermion) XX-staggered chain has many local integrals of motion and retains free-fermion integrability even by adding {stochastic noise due to quantum measurements} in the QSD protocol~\cite{DeLuca2019}, thus no random-state value is recovered. On the other hand, the SYK model has no local integrals of motion (the Hamiltonian is nonlocal) and the SRE relaxes to the random-value also in the unitary dynamics.} 
From the scaling with the system size of the fitting parameters{, and from the plots of the steady-state SRE versus the system size,} we conclude that the {{steady-state}} SRE is always linear in $L$, suggesting {that, for increasing system sizes, more $T$ gates (the ones not included in the set of Clifford gates) are needed to prepare the state resulting from the evolution under measurements.}

Perspectives of future research include applying the methods of Ref.~\cite{collura2024quantummagicfermionicgaussian}, dealing with an efficient evaluation procedure for the SRE in systems of noninteracting fermions, to describe the free-fermion XX-staggered chain for larger system sizes. Beyond that, one could apply to the nonintegrable cases the sampling scheme based on matrix product states described in Ref.~\cite{lami2023}. {These would be ways to corroborate the generalized Lorentzian dependence for larger system sizes. Furthermore, at least in the unitary case where no post-selection problem exists, one can think to experimentally measure the SRE using the method explained in Ref.~\cite{magmeasured}.} Finally it would be interesting to consider other monitored quantum systems, with the ultimate purpose to find a measurement transition in the behavior of the {{steady-state}} SRE akin to the entanglement transitions. 

{\it Note added in proofreading.} During the completion of this work we became aware of Refs.~\cite{Schiro,Hamma}, where the {nonstabilizerness} of the unitary dynamics and of the eigenstates of the SYK model is discussed.

\acknowledgments
We acknowledge M.~Collura, F.~Leiber, and X.~Turkeshi for useful comments on the manuscript and E.~Tirrito for fruitful discussions. G.\,P. and A.\,R. acknowledge financial support from PNRR MUR Project PE0000023-NQSTI. We acknowledge computational resources from the CINECA award under the ISCRA initiative, and from MUR, PON “Ricerca e Innovazione 2014-2020”, under Grant No.~PIR01\_00011 - (I.Bi.S.Co.). This work was supported by PNRR MUR project~PE0000023 - NQSTI, by the European Union’s Horizon 2020 research and innovation programme under Grant Agreement No~101017733, by the MUR project~CN\_00000013-ICSC (P.\,L.),  and by the  QuantERA II Programme STAQS project that has received funding from the European Union’s Horizon 2020 research and innovation programme under Grant Agreement No~101017733 (P.\,L.).
\appendix
\section{Numerical evaluation of the nonstabilizerness}\label{nonstabilizerness:sec}
To evaluate the nonstabilizerness, quantified through the stabilizer R\'enyi entropy (SRE)~\cite{leone2022,oliviero2022}, one must compute all expectations of the Pauli strings on the state. {Efficient protocols based on Monte Carlo sampling of the Pauli strings~\cite{lami2023,ding2025evaluatingmanybodystabilizerrenyi} and others specifically tailored for fermionic Gaussian states~\cite{collura2024quantummagicfermionicgaussian} already exist. In contrast, in our case we sample over all the Pauli strings, and we simplify the evaluation of the expectation of each one of them. Our protocol is best suited for the case where the dynamics is restricted to a subspace. To be concrete, we consider a set of qubits and restrict to the subspace with a zero total magnetization along the $z$ axis.}

We consider the set $\mathcal{P}_L$ of the Pauli strings
\begin{equation}\label{pipo:eqn}
\hat{P}=\prod_{j=1}^L\hat{\sigma}_j^{\alpha_j}\,,
\end{equation}
with $\alpha_j=0,\,1,\,2,\,3$ ($\hat{\sigma}_j^0 \equiv \hat{\mathbb{I}}$, $\hat{\sigma}_j^1 \equiv \hat{\sigma}_j^x$, $\hat{\sigma}_j^2 \equiv \hat{\sigma}_j^y$, $\hat{\sigma}_j^3 \equiv \hat{\sigma}_j^z$, and $\hat{\sigma}_j^{(x,y,z)}$ are the spin-1/2 Pauli operators). The SRE is defined as~\cite{leone2022, oliviero2022}
\begin{equation}\label{orco:eqn}
  \mathcal{M}_2(t) = -\ln \bigg[ \frac{1}{2^L}\sum_{\hat{P}\in\mathcal{P}_L}\braket{\psi_t|\hat{P}|\psi_t}^4\bigg]\,.
\end{equation}
To evaluate the expectations
%
 $A_P = \braket{\psi_t|\hat{P}|\psi_t}$,
%
we expand the state of the system $\ket{\psi_t}$ at time $t$ on the basis of the simultaneous eigenstates of $\hat{\sigma}_j^z$ (the so-called computational basis):
\begin{equation}
  \ket{\psi_t} = \sum_{\{s_j\}}C_{\{s_j\}} \ket{\{s_j\}}\,,
\end{equation}
where we restrict to the $\mathcal{N}_L=\binom{L}{L/2}$ configurations $\{s_j\}$ such that $\sum_{j=1}^Ls_j = 0$. (Dynamics is restricted to the subspace with vanishing total $z$ magnetization.) 
We can write the Pauli expectation as
%
  $A_P = \sum_{\{s_j\}\,,\,\{s_j'\}}C_{\{s_j\}}^*C_{\{s_j'\}}\braket{\{s_j\}|\hat{P}|\{s_j'\}}$.
%
Because both the states $\ket{\{s_j\}}$ and the operators $\hat{P}$ have a product structure, we can write
\begin{equation}
  \braket{\{s_j\}|\hat{P}|\{s_j'\}} = \prod_{j=1}^L\braket{s_j|\hat{\sigma}_j^{\alpha_j}|s_j'}\,.
\end{equation}
The elements of this product can be easily evaluated as
\begin{align}
  \braket{s_j|\hat{\sigma}_j^{\alpha_j}|s_j'} =&\; \, \delta_{\alpha_j,0}\,\delta_{s_j,s_j'} + \delta_{\alpha_j,1}(\delta_{s_j,s_j'+1}+\delta_{s_j,s_j'-1}) \nonumber\\
  &+ \delta_{\alpha_j,2}(-i\delta_{s_j,s_j'+1}+i\delta_{s_j,s_j'-1})\nonumber\\
  & + \delta_{\alpha_j,3}\,\delta_{s_j,s_j'}(\delta_{s_j,1}-\delta_{s_j,0})\,.
\end{align}
Given a configuration $\{s_j\}$ and a Pauli operator $\hat{P}$, there exists one and only one configuration $\{s_j'\}$ such that $\braket{s_j|\hat{\sigma}_j^{\alpha_j}|s_j'}\neq 0$, which can be easily constructed as follows. Given a site $j$, if the operator $\hat{P}$ in Eq.~\eqref{pipo:eqn} is such that $\alpha_j=0$ or $\alpha_j=3$, then $s_j'=s_j$, otherwise $s_j'=-s_j$. So the expectation of the Pauli operator $\hat{P}$ on the state $\ket{\psi_t}$ is given by
\begin{equation}
  \braket{\psi_t|\hat{P}|\psi_t} = \sum_{\{s_j\}}\braket{\{s_j\}|\hat{P}|\{s_j'\}}C_{\{s_j\}}^*C_{\{s_j'\}}\,,
\end{equation}
where the sum is restricted to the configurations $\{s_j\}$ such that $\sum_{j=1}^Ls_j = \sum_{j=1}^Ls_j' = 0$. {Notice that, although not explicitly written, $\{s_j'\}$ is a function of $\hat{P}$ and $\{s_j\}$ and varies along the summation.}

{We recall that}, in this way, we do not need to store the Pauli operators in the memory, and for each operator we only need to perform a single sum over {$\mathcal{N}_L=\binom{L}{L/2}$} terms, instead of evaluating the expectation of a $2^L\times 2^L$ matrix. {This approach allows simulation of systems up to $L = 14$ sites.}
%
%
%
\end{document}